# High-Performance Flexible Nanoscale Field-Effect Transistors Based on Transition Metal Dichalcogenides

Alwin Daus[1], Sam Vaziri[1], Victoria Chen[1], Cagil Koroglu[1], Ryan W. Grady[1], Connor S. Bailey[1], Hye Ryoung Lee[2], Kevin Brenner[1], Kirstin Schauble[1] and Eric Pop[1,3,*]

[1]Department of Electrical Engineering, Stanford University, Stanford, CA 94305, U.S.A.

[2]Geballe Laboratory for Advanced Materials, Stanford University, Stanford, CA 94305, U.S.A.

[3]Department of Materials Science & Engineering, Stanford University, Stanford, CA 94305, U.S.A.

[*]Corresponding author email: epop@stanford.edu

**Two-dimensional (2D) semiconducting transition metal dichalcogenides (TMDs) are good candidates for high-performance flexible electronics. However, most demonstrations of such flexible field-effect transistors (FETs) to date have been on the micron scale, not benefitting from the short-channel advantages of 2D-TMDs. Here, we demonstrate flexible monolayer $MoS_2$ FETs with the shortest channels reported to date (down to 50 nm) and remarkably high on-current (up to 470 $\mu A\ \mu m^{-1}$ at 1 V drain-to-source voltage) which is comparable to flexible graphene or crystalline silicon FETs. This is achieved using a new transfer method wherein contacts are initially patterned on the rigid TMD growth substrate with nanoscale lithography, then coated with a polyimide (PI) film which becomes the flexible substrate after release, with the contacts and TMD. We also apply this transfer process to other TMDs, reporting the first flexible FETs with $MoSe_2$ and record on-current for flexible $WSe_2$ FETs. These achievements push 2D semiconductors closer to a technology for low-power and high-performance flexible electronics.**

For several years, the "Internet-of-Things" (IoT) has been increasingly prevalent in the forecast of future electronics. From monitoring the environment and machines around us to the human body, IoT envisions electronics physically present in every aspect of our daily lives. While some devices may be realized on rigid silicon, there is a need for electronics with new non-planar form factors[1,2], which are thin and light, and can be conformally attached to objects with unusual shapes, on the human skin, or even implanted into the human body[1]. With these applications in mind, we need to realize electronics on flexible substrates that are robust to mechanical strain, easy to integrate, and capable of low-power consumption and high performance at the nanoscale[2,3].

Recent studies have suggested that 2D materials are good candidates for flexible substrates, because of their lack of dangling bonds, good carrier mobility in atomically thin (sub-1 nm) layers, reduced short-channel effects, and easy transfer onto arbitrary substrates[2,4-7]. Among these candidates, mono-layer transition metal dichalcogenides (TMDs) like $MoS_2$ are well-suited for low-power applications due to their good electronic band gaps (~2 eV)[8,9] which enable low off-currents (~fA $\mu m^{-1}$)[10,11]. How-ever, the performance of flexible TMD transistors with nanoscale channel lengths is still elusive be-cause of the difficulty of realizing such dimensions on flexible substrates[12], and due to TMD transfer processes which leave contamination or cause damage to the atomically-thin material[13-15]. The shortest flexible $MoS_2$ transistors (~68 nm) reported to date used 3-layer exfoliated material with on-currents of 135 $\mu A \mu m^{-1}$, ostensibly limited by their contact resistance[16]. For large-scale practical applications, $MoS_2$ must be synthesized by chemical vapor deposition (CVD), and the shortest channel (gate) re-ported was ~750 nm (~500 nm) with 85 $\mu A \mu m^{-1}$ on-current, also contact-limited[17]. There have been only a limited number of reports on flexible TMD field-effect transistors (FETs) other than $MoS_2$, including flexible $WSe_2$ FETs[18-21] and, to our knowledge, none on flexible $MoSe_2$ FETs to date.

In this work, we demonstrate flexible monolayer $MoS_2$ transistors with on-currents up to 470 $\mu A \mu m^{-1}$ at $V_{DS} = 1$ V in sub-100 nm channels, which is the highest reported to date for flexible $MoS_2$ FETs. These devices are achieved with a new transfer process for TMDs onto flexible substrates, including lithographically predefined metal contacts. The TMD is grown by high-quality CVD on a $SiO_2$/Si sub-strate and the critical contact separation is defined while the channel is still on the rigid substrate, enabling the nanoscale devices. Flexible polyimide (PI) is spin-coated onto the pre-patterned structures and all are released together, with the remaining process continuing on the PI. This approach enables record-achieving flexible FETs with $MoS_2$ and $WSe_2$, and the first flexible $MoSe_2$ transistors, all in the staggered configuration[22], i.e., with the channel sandwiched between source/drain and gate.

**Transfer Process with Embedded Contacts**

The TMDs were grown on $SiO_2$/Si substrates using chemical vapor deposition (CVD) as previously reported[23-26]. Subsequently, we lithographically patterned Au metal contacts on top. We chose Au be-cause of its good contact resistance[27] ($R_C$) to $MoS_2$ and its low adhesion[28] to $SiO_2$. Hence, both Au and the TMD (lacking out-of-plane dangling bonds) can be released from $SiO_2$ surfaces without damage, as shown below. After the definition of the metal contacts, we conformally cover the pre-patterned structures with ~5 $\mu m$ thick PI which is released together with the TMD and metal from the $SiO_2$/Si growth substrate, by immersion and agitation in DI water (additional details are provided in Methods).

Fig. 1a shows the processed $SiO_2$/Si substrate with TMD, contacts, and PI, Fig. 1b displays the release schematic, and Fig. 1c shows the transparent PI substrate after release. We note that a similar damage-free transfer of MoS$_2$ layers (without contacts) for coplanar[22] micron-sized FETs has recently been demonstrated over 4-inch wafers[29], indicating this approach could be scaled up.

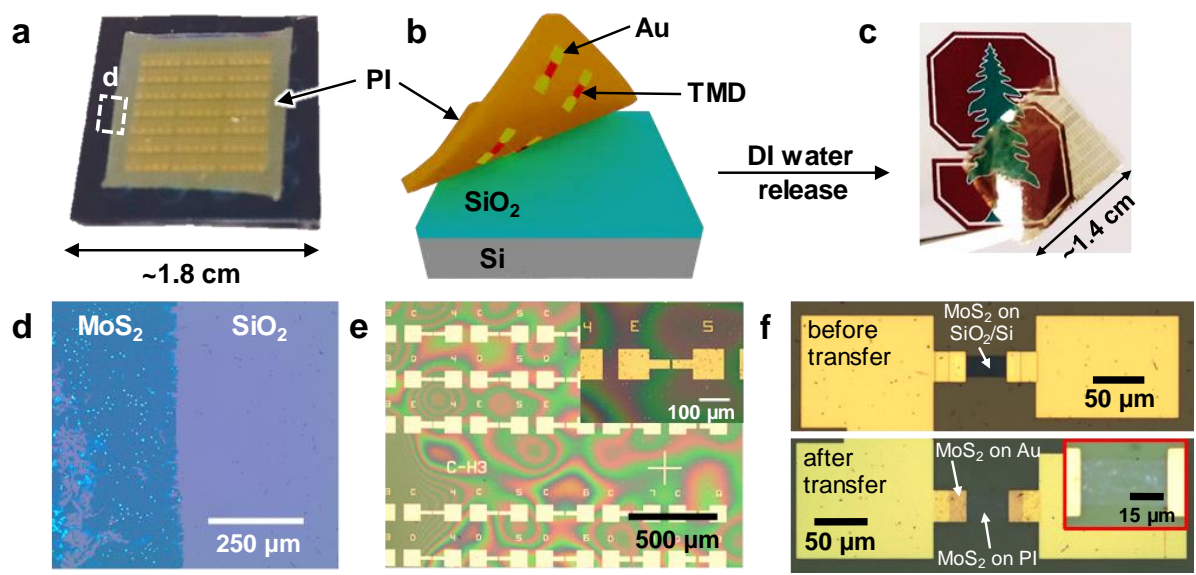

**Fig. 1 | Transfer process for 2D monolayers with contacts. a,** Optical image of MoS$_2$ and patterned metal covered by polyimide (PI) on SiO$_2$/Si. **b,** Schematic of the transfer process: the PI with embedded metal contacts and monolayer TMD are released from the rigid growth substrate. **c,** Optical image of the flexible PI after transfer. Optical microscope images of **d,** the SiO$_2$/Si growth substrate after transfer (with the bare SiO$_2$ surface where MoS$_2$ had previously been covered by PI), **e,** PI film with contacts and unpatterned MoS$_2$ after transfer and **f,** contacts with patterned MoS$_2$ on SiO$_2$/Si (top) and PI (bottom). Raman and photoluminescence (PL) spectra of MoS$_2$, WSe$_2$, MoSe$_2$ before and after transfer are shown in Supplementary Section B.

The microscope image in Fig. 1d shows that the TMD (here MoS$_2$) is completely delaminated from the area that had been covered by PI on the SiO$_2$/Si substrate. We can transfer continuous (Fig. 1e) as well as pre-patterned (Fig. 1f) MoS$_2$ films with embedded contacts, enabling us to realize devices based on several fabrication approaches. As shown later, we fabricated FETs with MoS$_2$, MoSe$_2$ and WSe$_2$, where only the contacts were patterned before transfer, minimizing process steps on unprotected TMDs (Fig. 1e and Supplementary Fig. S1), but leading to channel widths greater than the electrode widths (referred to as *Type A* devices). We also realized FETs where the MoS$_2$ channel was predefined by reactive ion etching (RIE) before transfer (Fig. 1f), which enables accurate channel width definition (referred to as *Type B*). Further details on device fabrication can be found in the Methods.

To confirm that the TMDs remain intact throughout the transfer process, we performed extensive Raman spectroscopy and photoluminescence (PL) measurements before and after transfer (Supplementary Fig. S2 and Section B). We observed that PI background signal and quenching on Au surfaces[30,31] affect the resolution and visibility of TMD peaks. The Raman and PL spectra on $SiO_2$/Si indicate TMDs with monolayer thickness, however, $MoSe_2$ also had regions with bilayers (see Supplementary Section B). Generally, the absence of major changes in Raman and PL before/after transfer indicates that mono- as well as multilayer TMDs can be readily transferred with this technique, without apparent damage. The electrical results presented below further confirm the excellent viability of this transfer approach.

**Flexible Top-Gated Field-Effect Transistors**

After the transfer process, the source/drain contacts are now embedded in the PI substrate and the TMD semiconductor is on top. To prevent contamination of this exposed TMD surface, we deposit an $Al_2O_3$ gate dielectric immediately after the transfer process and prior to any other patterning steps. The fabrication process is finalized with the gate metal definition, leading to a staggered device geometry. For $MoS_2$ FETs of *Type A*, we employ RIE to pattern the channel and gate dielectric together after the gate metal deposition, and additional fabrication details are given in Methods. The device cross-section is schematically shown in Fig. 2a and Fig. 2b-d display optical images of the $WSe_2$, $MoSe_2$ and $MoS_2$ FETs. Measured transfer and output characteristics of micron-scale FETs with $WSe_2$, $MoSe_2$ (both *Type A*) and $MoS_2$ (*Type B*) are presented in Fig. 2e-j, respectively, all displaying *n*-type behavior.

The extracted device parameters for all TMDs are listed in Table 1. Threshold voltage $V_T$ and extrinsic field-effect mobility $\mu_{FE,ext}$ were estimated at the maximum transconductance ($g_m$), using the measured $Al_2O_3$ gate oxide capacitance ($C_{ox} = 0.21$ to $0.32$ $\mu F$ $cm^{-2}$) from the TMD FETs obtained in accumulation (see Supplementary Fig. S7)[32]. The 2 $\mu$m long monolayer $WSe_2$ FET exhibits a maximum on-current $I_D = 3.5 \pm 0.05$ $\mu A$ $\mu m^{-1}$ (the source of the error bars is explained below) at a drain-source voltage $V_{DS} = 1$ V, which is over twice larger than the highest previously reported for flexible $WSe_2$ (using bilayer exfoliated material)[18]. The 3 $\mu$m long $MoSe_2$ FET reaches $I_D = 4.2 \pm 0.34$ $\mu A$ $\mu m^{-1}$ at $V_{DS} = 4$ V, which is to our knowledge the first demonstration of flexible $MoSe_2$ FETs.

The mobility and width-normalized current of *Type A* devices are listed with error bars because in these the channel width was not patterned and they were subject to (some) current spreading effects, which we account for with numerical simulations (see Supplementary Section K). For example, the unpatterned hexagonal crystals for the selenide-based FETs can be seen in Fig. 2b,c. Their measured data are shown in plain current units ($\mu A$) in Fig. 2e,f and Fig. 2h,i, respectively, but the error bars are included when presenting their width-normalized current ($\mu A/\mu m$), e.g. in Table 1.

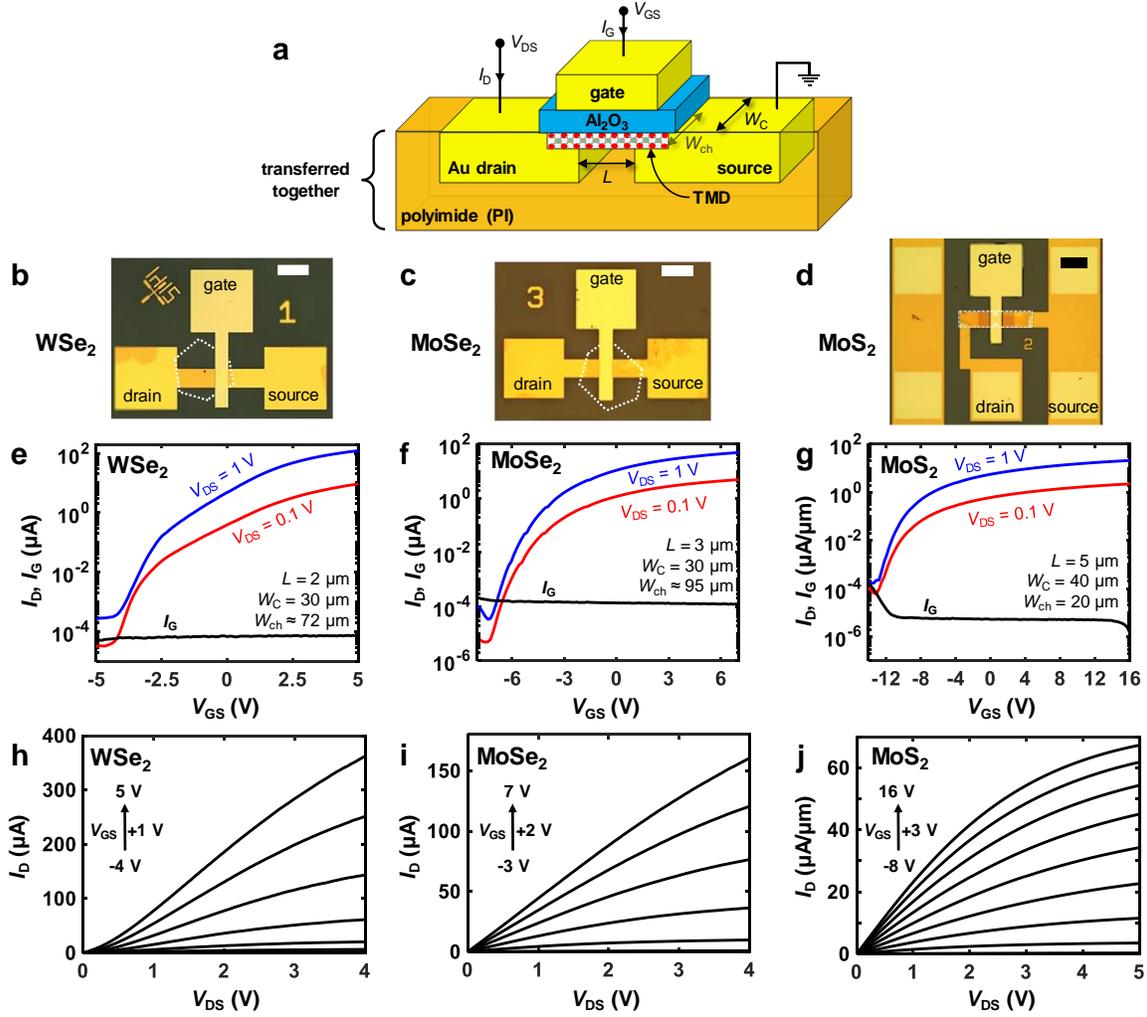

**Fig. 2 | Flexible field-effect transistors (FETs) with transition metal dichalcogenides (TMDs).**
**a,** Schematic cross-section. Optical microscope images of FETs with **b**, WSe$_2$ (*Type A*), **c**, MoSe$_2$ (*Type A*) and **d**, MoS$_2$ (*Type B*), scale bars: 50 µm. Measured transfer and output characteristics of **e, h**, WSe$_2$ (*Type A*), **f, i**, MoSe$_2$ (*Type A*) and **g, j**, MoS$_2$ (*Type B*). The gate current ($I_G$) is often negligible, although for some devices it can limit the on/off ratio.

The correction is not needed for our *Type B* devices because of the optimized geometry and modified fabrication process. Hence, Fig. 2g,j for *Type B* MoS$_2$ FETs are displayed in width-normalized units and the better quality of this material also enables larger $I_D \approx 67.3$ µA µm$^{-1}$ in a 5 µm long FET at $V_{DS}$ = 5 V. In comparison, *Type A* MoS$_2$ FETs had higher subthreshold swing SS and off-current, reducing their on/off ratio (Table 1 and Supplementary Fig. S8). Comparing device hysteresis[33], we find it ranges from ~0.1 V (WSe$_2$) to ~1.6 V (MoSe$_2$) for all devices and TMDs, indicating that the additional patterning step of *Type B* devices does not deteriorate TMD interfaces (see Supplementary Fig. S9). We have also verified the stability of the flexible TMD FETs under tensile bending and found negligible changes for a bending radius of 4 mm (Supplementary Fig. S10).

| Device | Channel Length (nm) | $\mu_{FE,ext}$ ($cm^2V^{-1}s^{-1}$) | $I_D$ at $V_{DS} = 1V$ ($\mu A\ \mu m^{-1}$) | $V_T$ (V) | SS (mV decade$^{-1}$) | on/off ratio |
|---|---|---|---|---|---|---|
| WSe$_2$ (*Type A*) | 2000 | 4.9 ± 0.07* | 3.5 ± 0.05* | 1.6 | 380 | $3 \times 10^5$ |
| MoSe$_2$ (*Type A*) | 3000 | 1.4 ± 0.09* | 1.3 ± 0.09* | -2 | 430 | $10^6$ |
| MoS$_2$ (*Type A*) | 5000 | 15 ± 2.2* | 5.4 ± 0.79* | 3.9 | 1700 | $3.6 \times 10^3$ |
| MoS$_2$ (*Type B*) | 5000 | 25.7 | 21 | -5.2 | 850 | $10^5$ |
| MoS$_2$ (*Type B*) | 100 | 7.2 | 229 | 0.6 | 730 | $2 \times 10^6$ |
| MoS$_2$ (*Type A*) | 70 | 20 ± 1.4* | 470 ± 45* | 6 | 1000 | $4 \times 10^3$ |

**Table 1 | Electrical parameters of flexible FETs.** The extrinsic field-effect mobility $\mu_{FE,ext}$ and threshold voltage $V_T$, were extracted from the maximum $g_m$ in the linear FET operating regime at a drain-source voltage $V_{DS} = 100$ mV. The subthreshold swing (SS) value denotes the extracted minimum. We note some $V_T$ variability, which is not unusual for 2D channels in academic fabrication facilities (also see Supplementary Fig. S15). * indicates values corrected for current spreading.

## Nanoscale MoS$_2$ Flexible Transistors

As MoS$_2$ has the most mature growth process with the highest electrical quality and best surface coverage, we further studied its FET scaling down to ~50 nm with electron-beam lithography (EBL) for source/drain contact patterning. Importantly, this nanoscale resolution is enabled by our approach wherein the contact patterning is first performed on the atomically smooth SiO$_2$/Si surface instead of the PI, which is prone to waviness, enhanced charging effects, and possible damage in EBL[34]. We also verified that this process is benign to MoS$_2$, performing Raman and PL measurements before and after EBL, finding no apparent evidence of damage to MoS$_2$ (Supplementary Section G). The remaining device fabrication and transfer were performed as described above.

Fig. 3a displays a top view optical image of a nanoscale channel after transfer, and a post-fabrication device cross-section. The cross-section reveals the Al$_2$O$_3$ gate dielectric covers the planar source/drain electrodes, including the ~100 nm nanogap between them, illustrating the absence of "steps" in surface topography enabled by this fabrication technique with contacts embedded in the flexible substrate. Electrical measurements of a similar *Type B* device with 100 nm long channel are shown in Fig. 3b,c revealing good on/off ratio ($> 10^6$), high $I_D \approx 303\ \mu A\ \mu m^{-1}$ (at $V_{DS} = 1.4$ V) and $\mu_{FE,ext} \approx 7.2\ cm^2V^{-1}s^{-1}$. The mobility appears smaller than in micron-scale devices due to greater contribution from contact resistance, as discussed below (for other device parameters see Table 1). Measured output characteristics (Fig. 3c) show signs of self-heating and velocity saturation[7,35] due to the onset of current saturation at lower $V_{DS}$ with higher gate-source voltages $V_{GS}$, which is similar to the self-heating of MoS$_2$ FETs on SiO$_2$/Si substrates. We estimate the temperature of this FET reaches ~172°C at the peak input power shown in Fig. 3c (see Supplementary Section H), and the Au contacts are primarily responsible for lateral heat spreading from the nanoscale device channel.

To gain additional insight into intrinsic device parameters, we extracted $I_D$ (at an overdrive $V_{ov} = V_{GS} - V_T = 8$ V) and $\mu_{FE,ext}$ for channel lengths from 50 nm to 10 μm in Fig. 3d,e. Measuring numerous devices allows us to comment both on "typical" and "best case" device performance. We use a model which relates $I_D$ and $\mu_{FE,ext}$ to $L$, $R_C$, and the intrinsic field-effect mobility $\mu_{FE}$. (The adapted model[36], the $V_T$ and $\mu_{FE,ext}$ extractions are described in Supplementary Sections C and J.) Fig. 3d,e reveals that $I_D$ plateaus and $\mu_{FE,ext}$ decreases at sub-1 μm channel lengths, which clearly indicates these devices are limited by $R_C$. The dashed black lines show the model for "typical" *Type B* devices (red circles) which we fitted with an average $\mu_{FE}$ (~12.9 cm²V⁻¹s⁻¹) for micron-scale devices where the impact of $R_C$ is small, and by setting $R_C = 5$ kΩ μm to follow the middle of the distribution for shorter $L$.

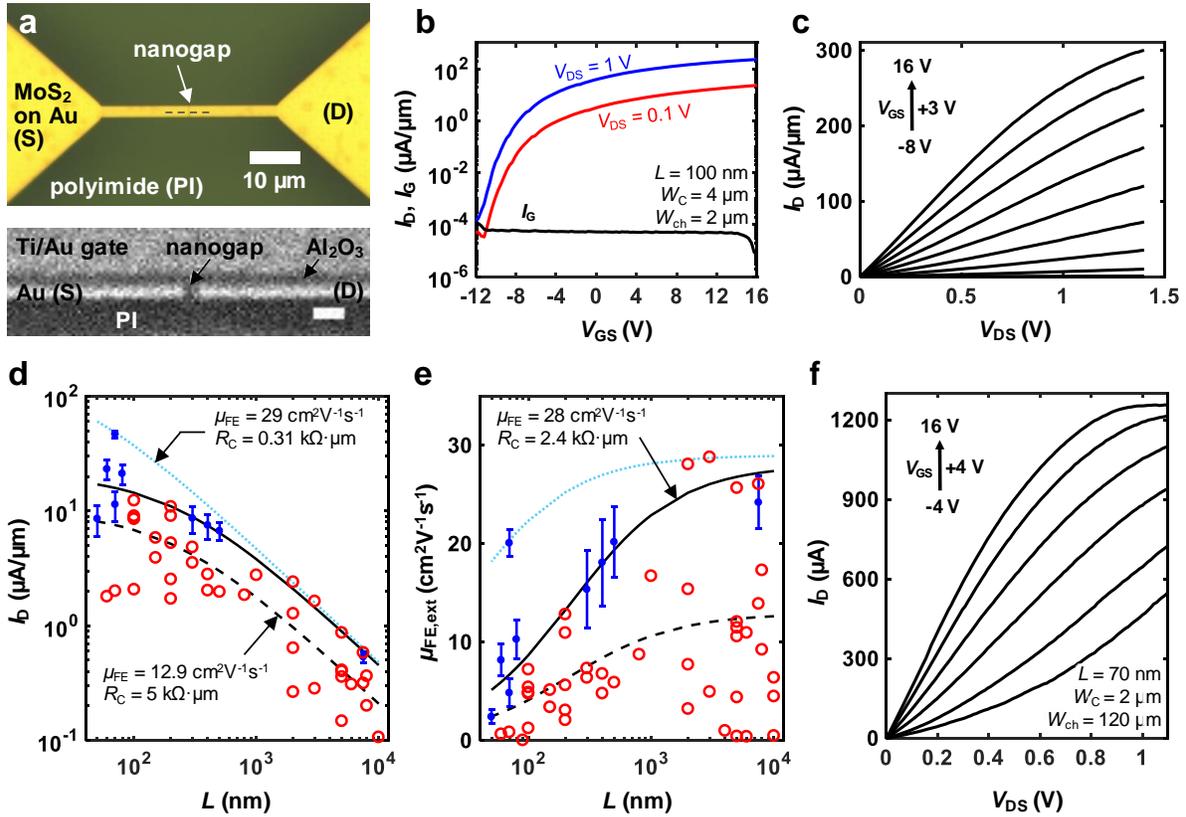

**Fig. 3 | Nanoscale MoS$_2$ field-effect transistors (FETs). a**, Optical microscope image of a nanoscale channel after transfer (top) and cross-section scanning electron microscope (SEM) image of a transistor with a channel length ($L$) = 100 nm (bottom, scale bar: 200 nm). The cross-section is taken at the dashed line in the top image. **b**, Transfer and **c**, output characteristics of a *Type B* MoS$_2$ FET with $L$ = 100 nm. **d**, Drain current $I_D$ and **e**, extrinsic field-effect mobility $\mu_{FE,ext}$ as a function of $L$ for ~50 devices at a drain-source voltage $V_{DS}$ = 0.1 V. $I_D$ is displayed at an overdrive voltage $V_{GS} - V_T$ = 8 V and $\mu_{FE,ext}$ was extracted at the maximum transconductance. Solid black and dashed black lines illustrate a fitted trend for "best" and "typical" *Type B* devices (red circles), while the blue dotted line combines best results for all FETs, including *Type A* devices (blue). **f**, Electrical characteristics of a MoS$_2$ FET (*Type A*) with $L$ = 70 nm showing remarkably high $I_D$ > 1.2 mA, which is ~470 μA μm$^{-1}$ after current spreading correction.

The solid black lines in Fig. 3d,e are based on a similar approach but using higher $\mu_{FE}$ (~28 cm²V⁻¹s⁻¹) to fit the best-performing *Type B* devices with $R_C = 2.4$ kΩ μm. Taking into account also "best" *Type A* devices (blue symbols and error bars, corrected for current spreading), we fit $R_C \approx 310$ Ω μm for one device (at $L = 70$ nm) and a slightly higher $\mu_{FE} = 29$ cm²V⁻¹s⁻¹, generating the blue dotted lines. The FET with highest on-current achieves an impressive $I_D = 470 \pm 45$ μA μm⁻¹ at $V_{DS} = 1$ V (see Supplementary Section I for electrical data and Supplementary Section K for current spreading correction), and its electrical characteristics are shown in Fig. 3f and Supplementary Fig. S14b. The presence of a "hero" device is not surprising when one hundred (or two hundred[24]) devices are measured, being both an indicator of academic fabrication variability, and of the promise of these 2D semiconductors if variability challenges are mitigated by industrial optimization. (We note that our *Type A* and *Type B* devices have similar variability, also see Supplementary Section I.)

Our estimated best-case $\mu_{FE}$ and $R_C$ are comparable to best reported values for monolayer $MoS_2$ on flexible substrates and on $SiO_2$/Si rigid substrates, respectively[17,24,35]. The highest on-current $I_D$ is over three times greater than in previous reports for flexible $MoS_2$ FETs[16], similar to the best TMD FETs on rigid substrates[37], and even comparable to flexible FETs based on graphene[38] and c-Si[39]. Moreover, this fabrication technique enables us to scale flexible $MoS_2$ FETs to the shortest channel lengths reported to date (Supplementary Fig. S14c).

Fig. 4 displays benchmarking of our flexible $MoS_2$ transistors compared to other technologies on flexible substrates. Displaying $\mu_{FE,ext}$ and $I_D$ (at $V_{DS} = 1$ V, unless noted otherwise) for flexible $MoS_2$ FETs *vs.* $L$ (Fig. 4a,b) reveals that nanoscale devices have received little attention until now (values listed in Supplementary Table S1)[4,16,17,29,40-49]. Fig. 4c compares the on-current and on/off ratio of the few existing sub-200 nm flexible FETs (at $V_{DS} = 0.5$ V, unless noted otherwise) revealing good performance of our $MoS_2$ even next to high-mobility materials (values listed in Supplementary Table S2)[39,50-54]. The on/off ratio of $MoS_2$ is many orders of magnitude higher than graphene, making $MoS_2$ FETs much more suitable for low-power applications among 2D channel materials. Compared to flexible c-Si FETs, flexible TMD FETs have a fundamentally different structure, with a sub-nanometer thin channel without out-of-plane dangling bonds. This enables shorter channel lengths, better mechanical robustness, and potentially lower cost (due to simple transfer processes), all advantageous for higher-performance and lower-power operation on flexible substrates.

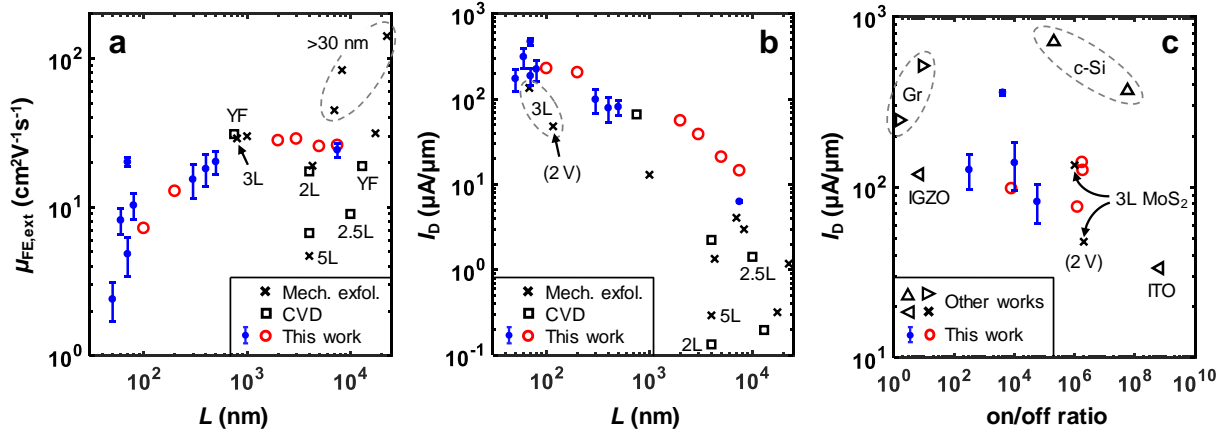

**Fig. 4 | Benchmarking flexible FETs. a**, Extrinsic field-effect mobility $\mu_{FE,ext}$ and **b**, drain current $I_D$ (at $V_{DS} = 1$ V) for flexible $MoS_2$ transistors as a function of channel length $(L)$[4,16,17,29,40-49]. Three studies reported $\mu_{FE}$ excluding contact resistance with the $y$-function method ("YF"). A few points correspond to 3-layer (3L) $MoS_2$, one of them at $V_{DS} = 2$ V, one $(L = 68$ nm) at unspecified voltage. Most CVD-grown $MoS_2$ are monolayers (1L), including this work, the other thicknesses are as labeled up to 5-layers (5L), and unlabeled points are all thicker exfoliated channels. All thicknesses and more details are listed in Supplementary Section L. **c**, Reported $I_D$ (at $V_{DS} = 0.5$ V) vs. on/off ratio for flexible FETs[39,50-54] with channel $L \leq 200$ nm. For comparison, two data points are shown for graphene ("Gr"), two for crystalline silicon ("c-Si"), two for oxide semiconductors (ITO and IGZO), and the others are for $MoS_2$ (our CVD monolayer data and two reports on 3L exfoliated).

## Conclusion

In summary, we demonstrated high-performance $MoS_2$ transistors on flexible substrates, with a novel transfer process which enabled making the shortest channels to date (~50 nm). We reached drive currents up to 470 μA μm⁻¹ ($L = 70$ nm, at $V_{DS} = 1$ V), among the highest for any monolayer 2D semiconductor, including those on rigid $SiO_2/Si$ substrates, and comparable to those of flexible graphene and c-Si transistors. The high current is achieved despite the low thermal conductivity of the PI substrate, as the short channel devices benefit from heat spreading by the Au contacts. We also applied the fabrication technique to other 2D semiconductors, reporting the first flexible $MoSe_2$ FETs and the highest current in flexible $WSe_2$ FETs. This fabrication approach could become a template for making flexible transistors with other materials where few short-channel demonstrations exist (e.g., oxides, organics, or carbon nanotubes). Together with further optimization of electrostatic control (thinner gate dielectrics or double-gates) and reduced parasitics (e.g., lower $R_C$) these results provide a promising outlook for incorporating flexible TMD electronics in low-power and high-performance IoT applications.

## Methods

Raman and Photoluminescence measurements

The Raman and PL measurements were performed on a HORIBA Scientific LabRAM HR Evolution spectrometer using an excitation wavelength of 532 nm. For Raman measurements on SiO$_2$/Si, an acquisition time, accumulations, laser power and optical grating of 5 s, 3, 0.14 mW, 1800 gr/mm were used, and the spot size is less than 1 μm. For the Raman measurements after transfer on PI or Au/PI surfaces, the acquisition time was increased to 45 s, while the other parameters remained the same. For PL measurements on SiO$_2$/Si, PI and Au/PI, an acquisition time, accumulations, laser power and optical grating of 5 s, 3, 0.14 mW, 600 gr/mm were used.

Device fabrication including transfer process

**Fabrication of device *Type A*:** The TMDs were grown on Si/SiO$_2$ substrates involving solid-source chemical vapor deposition (CVD) as previously reported[23-26]. The device fabrication was performed in the Stanford Nanofabrication Facility and Stanford Nano Shared Facilities. First, a 45 nm thick Au source/drain contact metal was deposited and patterned by electron-beam evaporation and lift-off. Optionally, the adhesion of the metal contacts to the later spin-coated PI layer can be improved by evaporating an additional Ti layer on top of Au prior to the lift-off, but it is not required. The lithographic patterning for that step was done via optical lithography (Heidelberg MLA 150 direct write lithography tool) for micron scale channel length, and via EBL for sub-micron scale channel lengths. The EBL parameters can be found below. Then, ~5 μm thick PI layer (PI-2610, HD MicroSystems) was spin-coated on top, baked at 90$^o$C and 150$^o$C on a hotplate for each 90 s, and finally cured in a nitrogen oven at 250ºC for 30 minutes. Prior to PI spin-coating, the outside edges of the silicon substrate were protected with tape to facilitate the release of PI from the Si substrate (Fig. 1a). The transfer was performed in DI water by initially mechanically releasing the outside edges with a tweezer followed by agitation until the PI substrate together with the metal contacts and TMD was floating on the DI water surface. After nitrogen blow-drying the substrate, 1.5 nm thick Al blanket film was deposited on top by electron-beam evaporation. This film acts as a seed layer for the subsequent atomic-layer deposition (ALD) of an $Al_2O_3$ gate dielectric at 200°C. Note, we used 35 nm $Al_2O_3$ for the $MoS_2$ devices and 23 nm thick $Al_2O_3$ for $MoSe_2$ and $WSe_2$ devices. This yields $C_{ox} \approx 0.21$ to $0.32$ $\mu F\,cm^{-2}$, directly measured in Supplementary Fig. S7. The oxide thicknesses were chosen to ensure higher device yield and to have numerous FETs for measurement. After the ALD, the gate metal was deposited by electron-beam evaporation of Ti/Au (5/60 nm) and patterned by optical lithography and lift-off. This concluded the fabrication for $MoSe_2$ and $WSe_2$ devices. For $MoS_2$ devices, as a final step the $Al_2O_3$ and $MoS_2$ were patterned together using reactive ion etching (Oxford 80 RIE) in $CF_4{:}O_2$ at gas flows of 50 sccm:5 sccm, 150 W power and a pressure of 30 mTorr.

**Fabrication of device *Type B*:** Until the source/drain contact metallization, the fabrication of *Type B* devices was the same as for type A. However, after source/drain metallization the $MoS_2$ channels were patterned by reactive ion etching (Oxford 80 RIE) in $CF_4{:}O_2$ at gas flows of 50 sccm:10 sccm, 100 W power and 30 mTorr pressure, followed by in situ surface cleaning with $O_2$ plasma (20 W, 10 mTorr, 40 sccm). Then, contact pads and leads were defined by optical lithography, e-beam evaporation of Au/Ti (60 nm/5 nm) and lift-off. This was followed by PI spin-coating, curing and transfer of all structures in the same way as described for device *Type A*. After the gate dielectric deposition (same as for *Type A* devices), via holes for probing source/drain electrodes were wet etched in Al etchant at 40°C. Finally, the gate metallization was done similarly as for device *Type A*.

Electron-beam lithography (EBL) on $MoS_2$

We used a double layer of poly(methyl methacrylate) (PMMA) for lift-off patterns defined by EBL. The bottom and top layer were 50 nm thick 495K A2 PMMA and 200 nm 950K A4 PMMA, respectively. EBL was performed on a JEOL JBX 6300 lithography system at a dose of 900 $\mu C\,cm^{-2}$ and an acceleration voltage of 100 kV.

## Electrical measurements

All transistors were tested with a Keithley 4200 on a probe station in ambient air. For the bending experiments, the substrates were attached to a metallic cylindrical rod with a radius of 4 mm.

## Acknowledgement

A.D. is in part supported by the Swiss National Science Foundation's Early Postdoc.Mobility fellowship (grant P2EZP2_181619) and in part by Beijing Institute of Collaborative Innovation (BICI). R.W.G., C.S.B. and K.S. acknowledge the National Science Foundation (NSF) Graduate Research Fellowship. K.S. also acknowledges the support of the Stanford Graduate Fellowship. All authors thank the Stanford Nanofabrication Facility and Stanford Nano Shared Facilities for enabling device fabrication and characterization, funded under NSF award ECCS-1542152. E.P. and S.V. acknowledge support from the Stanford SystemX Alliance.

## Author Contributions

A.D. conceived the work and performed the device fabrication and characterization. A.D. and S.V. developed the TMD transfer process. R.W.G. did the $MoS_2$ CVD growth and C.S.B. did the $WSe_2$ and $MoSe_2$ CVD growths. V.C. carried out the electron beam lithography. A.D. performed optical material analysis with help from K.S and K.B. H.R.L. did the FIB-SEM cross-sections. C.K. set up numerical current spreading simulations and thermal simulations with E.P. A.D. analyzed all data and wrote the manuscript with help from V.C., C.K, and E.P. All authors revised and commented on the manuscript. E.P. supervised the work.

## Competing Interests

The authors declare no competing interests.

# High-Performance Flexible Nanoscale Field-Effect Transistors Based on Transition Metal Dichalcogenides

Alwin Daus[1], Sam Vaziri[1], Victoria Chen[1], Cagil Koroglu[1], Ryan W. Grady[1], Connor S. Bailey[1], Hye Ryoung Lee[2], Kevin Brenner[1], Kirstin Schauble[1] and Eric Pop[1,3,*]

[1]Department of Electrical Engineering, Stanford University, Stanford, CA 94305, U.S.A.

[2]Geballe Laboratory for Advanced Materials, Stanford University, Stanford, CA 94305, U.S.A.

[3]Department of Materials Science & Engineering, Stanford University, Stanford, CA 94305, U.S.A.

[*]Corresponding author email: epop@stanford.edu

## A. Optical microscope images for WSe₂ and MoSe₂ before and after the transfer process

Supplementary Fig. S1a,c displays the hexagonally shaped WSe$_2$ and MoSe$_2$ crystal grains grown on SiO$_2$/Si substrates after the patterning of source/drain metal contacts, and *before* the transfer. As visible here, the Au contacts are on top of the TMDs. When the polyimide (PI) is applied on top, it uniformly

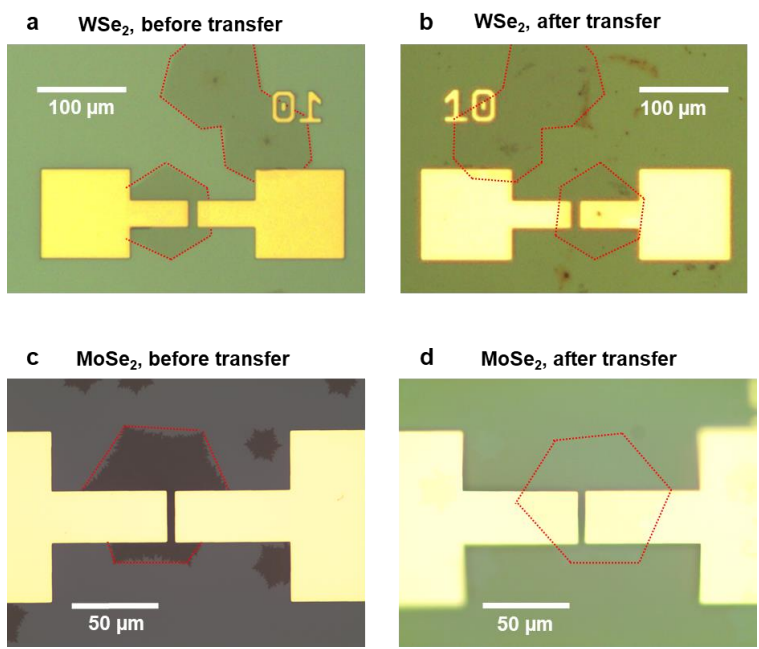

**Supplementary Fig. S1 | Microscope images of WSe₂ and MoSe₂ with patterned Au metal electrodes. a**, WSe$_2$ on SiO$_2$/Si before transfer. **b**, WSe$_2$ on polyimide (PI) after transfer. **c**, MoSe$_2$ on SiO$_2$/Si before transfer. **d**, MoSe$_2$ on PI after transfer.

covers and embeds the contacts. After releasing the PI together with metals and TMDs from the $SiO_2$/Si substrates (main text Fig. 1b), the substrate is flipped over as visible in the numbering ("10") when comparing Supplementary Fig. S1a,b. This also leads to a reversal of the material stack, where the TMDs are on top of Au/PI or PI, as visible in Supplementary Fig. S1b,d.

## B. Raman spectroscopy and Photoluminescence measurements before and after transfer process

The different TMDs were monitored with Raman spectroscopy and photoluminescence (PL) throughout the transfer process to investigate any changes in material properties. Supplementary Fig. S2a-f display the spectra for $MoS_2$, $WSe_2$ and $MoSe_2$ before and after transfer. Since we have deposited and patterned metal contacts before the transfer, released all materials together, and flipped the flexible PI substrate, we were able to measure the TMDs after transfer on the metal surface and on the PI surface.

We observed that the measurements directly on PI (without a metal in between the TMD and PI) have a broad background signal, which is absent on the $SiO_2$/Si substrates and on Au surfaces. This background signal is in the range where we expect the vibrational modes of the TMDs, and there is, for instance, a significant overlap with the PL energy maximum of monolayer $MoS_2$ (Supplementary Fig. S2b). The Raman and PL measurements for bare PI (on Si) are shown in Supplementary Fig. S2g,h for comparison. Because of this background signal, the $MoS_2$ Raman and PL peaks are buried and not visible in our measurements on PI. However, the peaks of $WSe_2$ and $MoSe_2$ on the PI surface can be resolved (Supplementary Fig. S2c-f). Further, the insertion of Au between PI and the TMD suppresses this background signal and enables the detection of the Raman signature of all three TMDs.

We find that the PL peaks for $WSe_2$ and $MoSe_2$ can be detected on PI and Au/PI despite the strong PL quenching that is known to appear on Au surfaces[30,31]. The PL peak energies of $MoS_2$, $WSe_2$ and $MoSe_2$ are ~1.86 eV, ~1.59 eV and ~1.54 eV, all indicating monolayer thickness[25,55-57]. While these results were consistent for $MoS_2$ and $WSe_2$ across the substrate, we found that $MoSe_2$ had areas with monolayers and bilayers (~50%) (Supplementary Fig. S3), where the PL peak is shifted towards ~1.50 eV and its intensity is significantly reduced. The noticeable spread in PL energies for $MoS_2$ can be attributed to a variety of effects such as nanoscale bilayer regions[23,24] or small local variations in strain or doping. For $MoSe_2$, however, we find two sets of PL energy peak positions, which indicate that some areas mainly consist of monolayers and some mainly of bilayers (~50% each)[25,55].

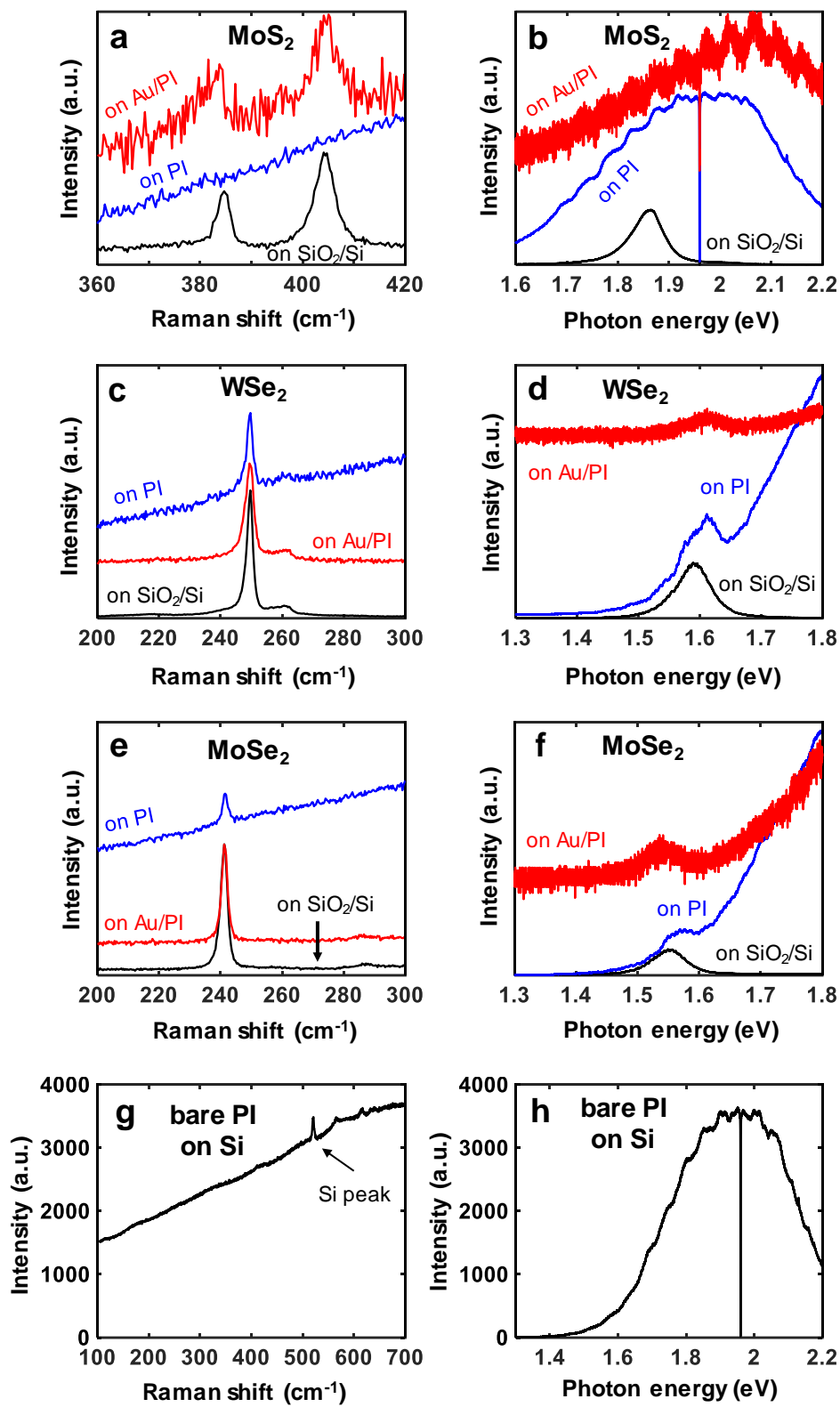

**Supplementary Fig. S2 | Raman (left column) and photoluminescence (right column) spectra.**
**a,b,** MoS$_2$. **c,d,** WSe$_2$. **e,f,** MoSe$_2$. **g,h** bare polyimide (PI) on silicon.

Changes in the Raman and PL spectra before vs. after transfer can be interpreted as strain release effects or phonon interactions with the bottom surfaces (PI or Au), as will be discussed in the following. For MoS$_2$ on Au, we find that the E' peak shifts by about -1.8 cm$^{-1}$ and its full-width-half-maximum (FWHM) increases, whereas the A$_1$' peak does not change discernably (see Supplementary Fig. S4), which has been observed for non-transferred Au/MoS$_2$ stacks and thus cannot be correlated with the transfer process. Possible mechanisms for this E' peak shift and broadening can be tensile strain induced from the Au deposition[58,59] or electron-phonon interactions due to Au plasmons[60,61]. For WSe$_2$, the changes in the Raman and PL spectra are small (Supplementary Fig. S5). The minor shifts in the PL peak position and Raman E' peak of about +0.02 eV and -0.1 to -0.2 cm$^{-1}$, respectively cannot be consistently correlated with any strain release during transfer[62], and may be related to small effects from interactions with the substrate[63]. It is difficult to deduce any strain effects from PL and Raman for MoSe$_2$, due to small changes and existence of mono- and bilayers adding uncertainty to the Raman and PL analysis. Still, the results suggest the possibility of slight strain changes in tensile and compressive directions on Au/PI and PI surfaces, respectively (see Supplementary Fig. S6)[64-66].

Overall, the FWHM of the Raman and PL peaks for all the TMDs do not increase except on Au electrodes, where previously discussed plasmonic effects could be the leading cause. This indicates that the disorder, which would be affected by crystal grain size or defect density, in the materials is not increased throughout the transfer process[67,68]. This conclusion is also supported by the good electrical properties, which are comparable on the materials after transfer with those before transfer (i.e. on rigid SiO$_2$/Si substrates) from previous studies using the same CVD material type[23,24,69].

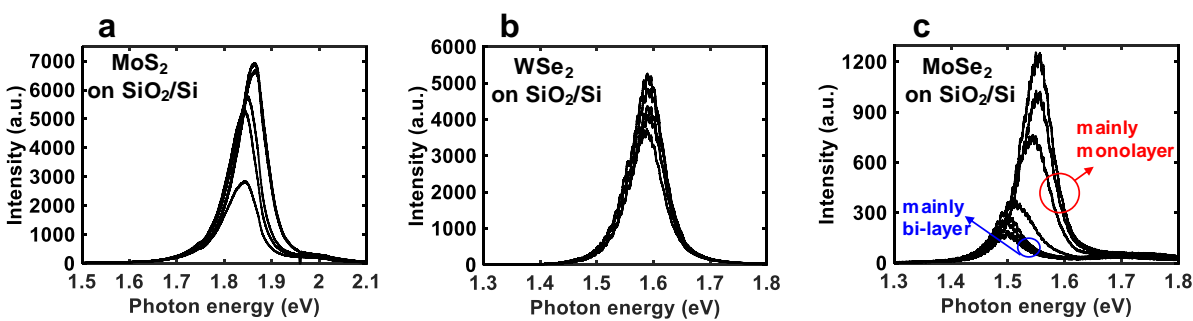

**Supplementary Fig. S3** | PL measurements before transfer of **a,** MoS$_2$, **b,** WSe$_2$ and **c,** MoSe$_2$.

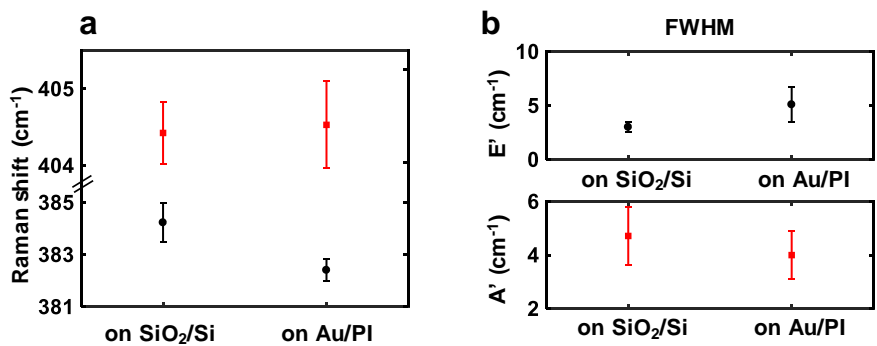

**Supplementary Fig. S4 |** Averaged Raman (over ~5 spots on the same chip) **a**, peak positions and **b**, Full-width-half-maximum (FWHM) of MoS$_2$ before transfer (as-grown, on SiO$_2$/Si substrate) and after transfer (on Au/PI).

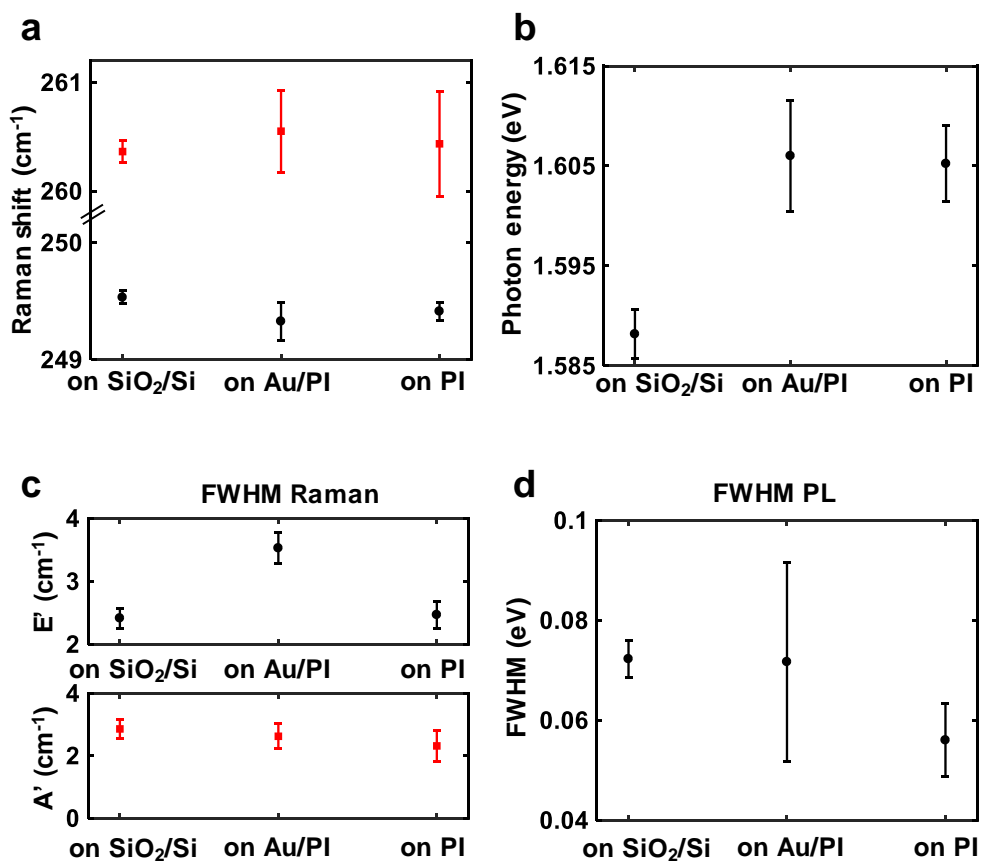

**Supplementary Fig. S5 |** Averaged (over ~5 spots on the same chip) **a**, Raman and **b**, photoluminescence (PL) peak positions of WSe$_2$ before and after transfer. Averaged full-width-half-maximum (FWHM) for **c,** the Raman and **d,** the PL measurements.

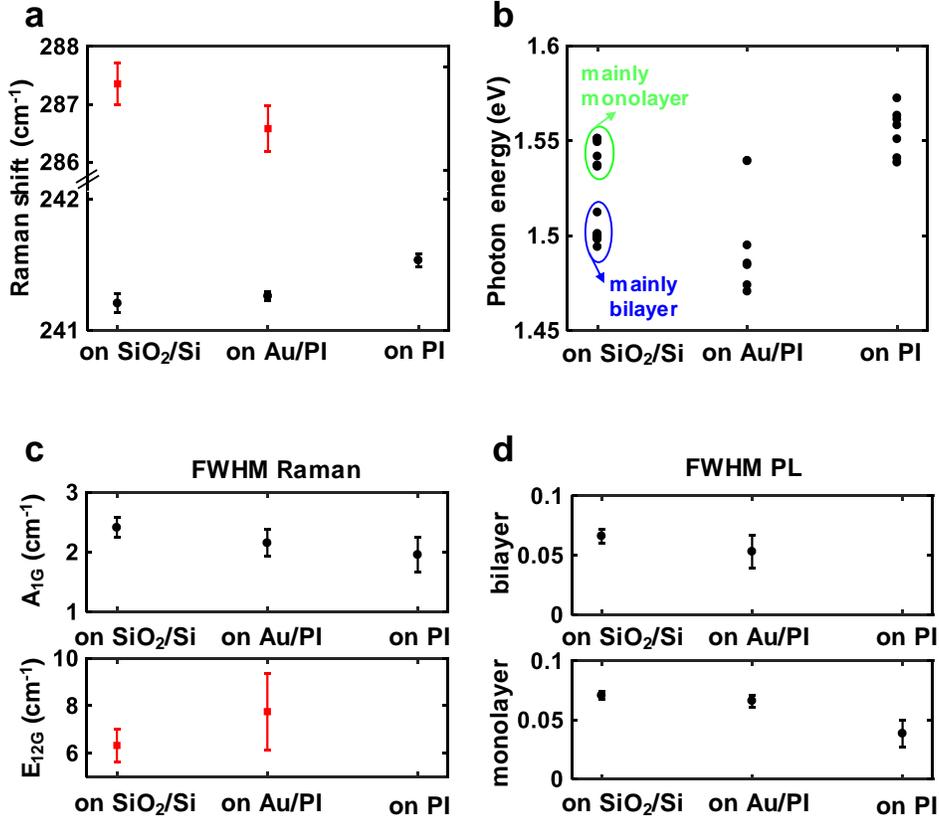

**Supplementary Fig. S6 | a**, Averaged Raman (over ~5 spots on the same chip) and **b,** photoluminescence (PL) peak positions of MoSe$_2$ before and after transfer. Averaged full-width-half-maximum (FWHM) for **c**, the Raman and **d**, the PL measurements (~10 spots across two chips). Note, missing data e.g., for the Raman E$_{12G}$ peak on PI means that these peaks could not be detected on that particular surface.

## C. Mobility, Gate Capacitance, and Threshold Voltage Extraction

We performed the extraction of the extrinsic field-effect mobility $\mu_{FE,ext}$ and threshold voltage $V_T$ from the $g_m$ maximum based on the following equation (valid for small drain-source voltages $V_{DS}$):

$$g_m = \frac{\partial I_D}{\partial V_{GS}} = \frac{\mu_{FE,ext} C_{ox} V_{DS} W}{L},$$

where the $I_D$ plotted vs. $V_{GS}$ can be fitted linearly to obtain $\mu_{FE,ext}$. Furthermore,[70,71] the intercept with the $V_{GS}$ axis yields $V_T$. The channel width $W$ and the channel length $L$ are given by the device geometry. The gate oxide capacitance per unit area ($C_{ox}$) is determined by connecting source and drain of the transistors to ground, and applying a voltage to the gate electrode. We then perform small-signal capacitance-voltage ($C$-$V$) measurements where the direct-current (dc) voltage is swept while applying an alternating-current (ac) voltage with an amplitude = 100 mV and frequency = 20 kHz. The results for Al$_2$O$_3$ gate dielectrics (including 1.5 nm Al seed layer) deposited in 300 cycles (MoS$_2$ FETs) and 200 cycles (WSe$_2$ and MoSe$_2$ FETs) are shown in Supplementary Fig. S7a and Fig. S7b, respectively.

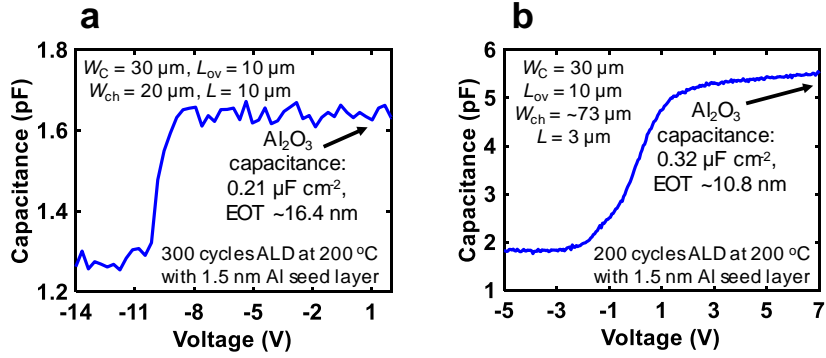

**Supplementary Fig. S7 | Measured capacitance-voltage (*C-V*) characteristics of flexible TMD-FETs.** Typical *C-V* for **a**, MoS$_2$ FETs and **b**, MoSe$_2$ and WSe$_2$ FETs, where the calculated Al$_2$O$_3$ capacitance density and equivalent oxide thickness (EOT) represent average values (~5 devices). *L*$_{ov}$ is the overlap length between the gate electrode and the source/drain electrodes.

As all transistors are *n*-channel devices, we estimate the capacitance of the Al$_2$O$_3$ gate dielectrics at positive bias voltage (when the channel is in accumulation) by dividing the measured capacitance (Supplementary Fig. S7) with the overlap area of the gate with the source, drain, and semiconductor channel[32]. For MoS$_2$ FETs the extracted $C_{ox} \approx 0.21$ µF cm$^{-2}$ or an equivalent oxide thickness (EOT) ~ 16.4 nm, and for WSe$_2$ and MoSe$_2$ FETs we obtain $C_{ox} \approx 0.32$ µF cm$^{-2}$ or EOT ~ 10.8 nm. Estimating the relative dielectric constant $\varepsilon_r$ from $C_{ox}$ and the Al$_2$O$_3$ thickness obtained by ellipsometry, we find $\varepsilon_r \approx$ 7-8, which is in the expected range[72,73]. Note, the ellipsometric thickness of 200 cycles and 300 cycles of atomic-layer deposited (200°C) Al$_2$O$_3$ measured on silicon is around 22 nm and 35 nm, respectively. However, for the transistor capacitance the 1.5 nm oxidized Al seed layer adds to the overall thickness, while the optical lithography and lift-off process of the top (gate) electrode exposes the Al$_2$O$_3$ to the basic photoresist developer which can etch the material, thus slightly reducing its thickness.

## D. Flexible MoS$_2$ Field-Effect Transistors of *Type A*

Supplementary Fig. S8 displays a top-down optical microscope image and the electrical characteristics of a flexible MoS$_2$ field-effect transistor (FET) of *Type A* with 5 µm channel length. The device exhibits an extrinsic field-effect mobility $\mu_{FE,ext}$ ~15 cm$^2$V$^{-1}$s$^{-1}$, on-current $I_D$ ~ 5.4 µA µm$^{-1}$ at a drain-source voltage $V_{DS} = 1$ V, threshold voltage $V_T = 3.9$ V, minimum subthreshold swing SS = ~1.7 V decade$^{-1}$ and on/off ratio ~3.6×10$^3$. Note that $\mu_{FE,ext}$ and $I_D$ have been corrected for current spreading, which is described in Section K below.

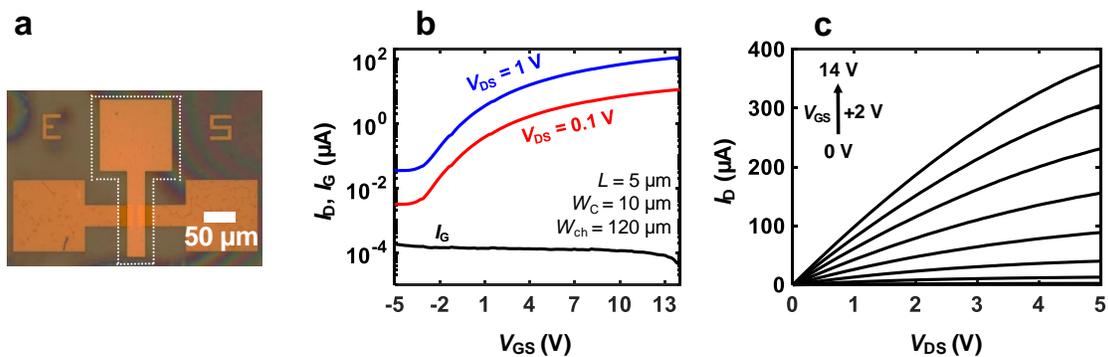

**Supplementary Fig. S8 | MoS₂ field-effect transistors (FETs) of *Type A*. a,** Top-down optical microscope image. **b,** Measured transfer characteristics. **c,** Output characteristics.

## E. Hysteresis

Comparing hysteresis[33] for *Type A* and *Type B* devices, we find similar maximum values of ~1.2 V for MoS₂ FETs, which indicates that the additional etch step before transfer does not deteriorate the TMD interfaces (Supplementary Fig. S9a,b). The WSe₂ FET displays low hysteresis ~0.1 V (Supplementary Fig. S9c), and the MoSe₂ FET has a maximum hysteresis of about 1.6 V (Supplementary Fig. S9d).

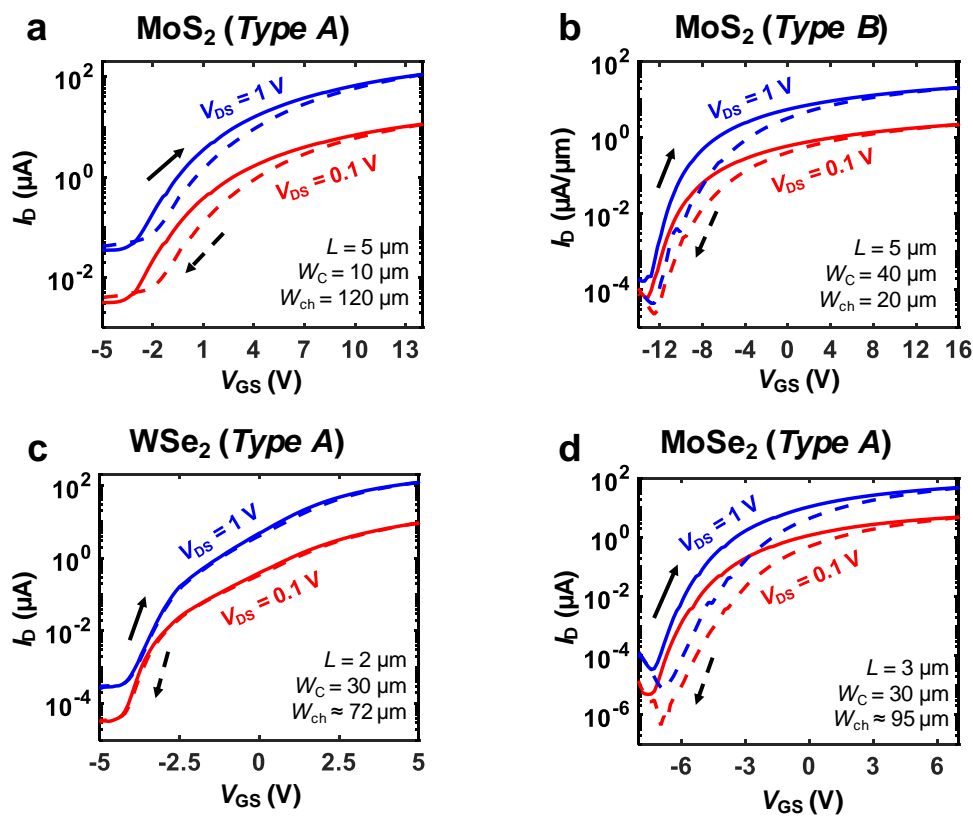

**Supplementary Fig. S9 | Hysteresis in flexible transition metal dichalcogenide field-effect transistors. a,** MoS₂ (*Type A*). **b,** MoS₂ (*Type B*). **c,** WSe₂ (*Type A*). **d,** MoSe₂ (*Type A*).

## F. Bending of flexible TMD FETs

Flexible electronics need to remain unaltered when mechanically deformed, for instance, by bending the substrate. While the ductility of materials matter for the maximum strain that flexible electronics can sustain[74], the easiest way to minimize impacts of strain on flexible electronic devices is to minimize the substrate thickness. The strain at a given bending radius can be approximated as[75]:

$$strain = \frac{d}{2r},$$

where $d$ is the substrate thickness and $r$ is the radius of curvature. Thus, by minimizing $d$ to a few micrometers, the strain at common bending radii on the order of millimeters is minimized. We show this by using a ~5 µm thick PI substrate and bending it to a radius of 4 mm, which results in ~0.063% of strain. Consequently, the electrical characteristics of the TMD FETs remain unaltered in this condition as shown in Supplementary Fig. S10.

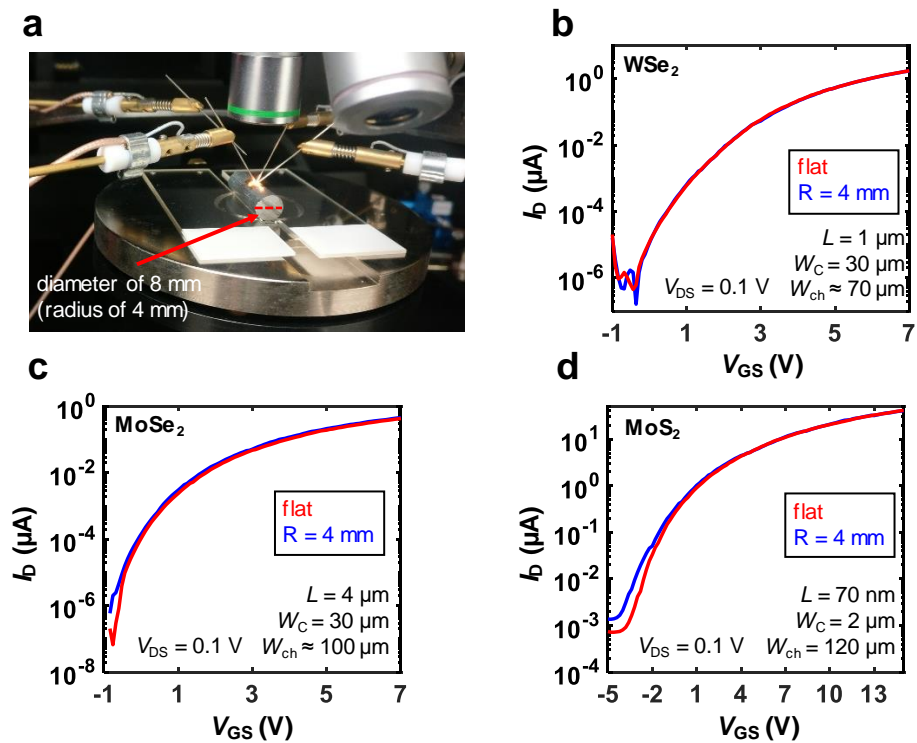

**Supplementary Fig. S10 | Mechanical bending of flexible TMD-FETs. a**, Photograph of the measurement setup where the FETs are bent to a tensile radius of 4 mm. Measured transfer characteristics of **b**, WSe$_2$ (*Type A*), **c**, MoSe$_2$ (*Type A*) and **d**, MoS$_2$ (*Type A*). All show effectively no dependence on substrate bending at the 4 mm radius.

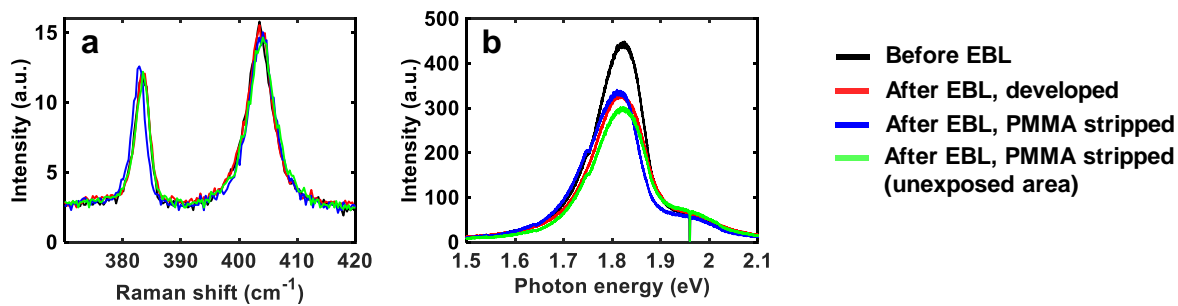

**Supplementary Fig. S11 | Optical material analysis of MoS₂ throughout an electron-beam lithography (EBL) process involving spin-coating and stripping of poly(methyl methacrylate) (PMMA). a**, Raman spectra. **b**, Photoluminescence spectra.

## G. Electron Beam Lithography on top of MoS₂

Previous reports have indicated that MoS₂ could be damaged by highly energetic electron beams which cause strain and defect formation[76-79]. We investigated this for our EBL process (details in the Methods section) performing Raman and PL measurements before and after the electron beam exposure and development of the poly(methyl methacrylate) (PMMA) layer that was used for the lift-off of the later deposited source/drain metal. Supplementary Fig. S11 displays exemplary Raman and PL spectra, while Supplementary Fig. S12 provides the analysis (averages over 8 spots) of Raman and PL peak center shifts as well as the changes in the intensity and FWHM of the PL spectra. We found negligible differences in the Raman spectra. The slight reduction in the PL intensity (Supplementary Fig. S12) occurs independently of the exposure to the electron beam, and may be caused by PMMA residues or minor effects from processing and aging of the material. Thus, we conclude that here we do not cause significant damage during the EBL process, which may have been due to our 200 nm thick high molecular weight (950K) PMMA layer on top that should reduce the impact energy and dosage of electrons that hit the $MoS_2$. The other parameters of our EBL process and the device fabrication can be found in the Methods section.

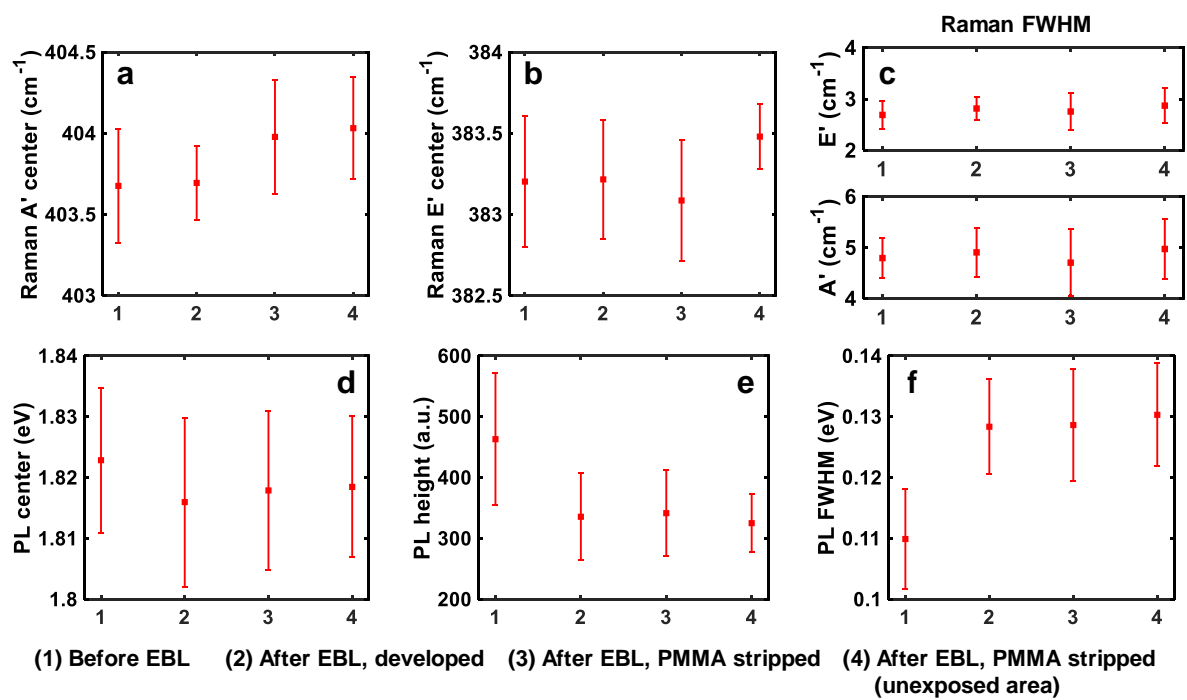

**Supplementary Fig. S12 | Averaged optical material analysis of MoS₂ during electron-beam lithography (EBL) with spin-coating and stripping of poly(methyl methacrylate) (PMMA). a**, Raman A' peak center. **b**, Raman E' peak center. **c**, Averaged full-width-half-maximum (FWHM) of the Raman peaks. **d**, Photoluminescence (PL) peak center. **e**, PL peak height. **f**, PL FWHM.

## H. Temperature Rise Estimates

We estimate the $MoS_2$ channel temperature rise in our nanoscale *Type B* devices ($W_{ch} < W_C$) using the thermal model in Supplementary Fig. S13 below. The heat sink at $T_0 \approx 20°C$ is below the PI film during measurements, and the PI film thickness is ~5 μm, larger than any device dimensions here. Three parallel thermal resistance pathways are of importance[80], direct heat spreading from the channel itself,

$$R_{ch} \approx \frac{1}{4k_{PI}(LW_{ch})^{1/2}}$$

and heat spreading by the Au contacts, $R_S = R_D \approx L_H/(k_C t_C W_{ch})$ where the thermal healing length

$$L_H \approx \left(\frac{k_C}{4k_{PI}} t_C \sqrt{W_{ch}}\right)^{2/3}.$$

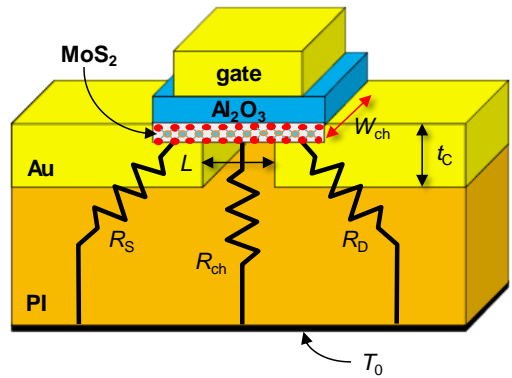

**Supplementary Fig. S13 | Thermal model for estimating the channel peak temperature.** There are three parallel thermal resistance paths from source ($R_S$), drain ($R_D$) and the channel ($R_{ch}$) to the heat sink on the bottom of polyimide (PI) with a temperature of $T_0 = 20°C$. $W_{ch}$, $L$ and $t_C$ represent the channel width, channel length, and Au contact thickness.

The peak temperature rise is estimated as $\Delta T \approx P'W_{ch}(1/R_{ch} + 2/R_S)^{-1}$, where $P' = 420$ W/m is the maximum power input achieved in Fig. 3c, and the relevant dimensions are $L = 100$ nm, $W_{ch} = 2$ µm, $t_C = 45$ nm. We take the thermal conductivity of the thin Au contact $k_C \approx 150$ $Wm^{-1}K^{-1}$, approximately $2\times$ lower[81] than that of bulk Au and the thermal conductivity of PI at elevated temperatures ($\geq 100°C$)[82] as $k_{PI} \approx 0.2$ $Wm^{-1}K^{-1}$. With these assumptions, we estimate $\Delta T \approx 152$ K or an operating temperature around 172°C, which is >90% dominated by heat spreading through the source/drain electrodes. This temperature may be slightly overestimated, because heat spreading through the top gate electrode was ignored. However, the thermal boundary resistances between the various materials have also been ignored in this simple model. From the manufacturer process guide[83], we note the PI glass transition temperature is 360°C and the decomposition temperature is 620°C. Thus, these devices have additional headroom to operate at higher power, thanks to heat spreading through their source/drain electrodes.

## I. Additional *Type A* Device Data and Overall Variability

The variability of $I_D$ for *Type A* devices shows a similar distribution like devices of *Type B* (compare Supplementary Fig. S14a with Fig. 3d). The transfer characteristic of the device with the highest on-current ($L = 70$ nm, from Fig. 3f) is displayed in Supplementary Fig. S14b. The shortest channel lengths which we realized were 50 nm and an exemplary electrical characteristic and scanning-electron microscopy cross-section are shown in Supplementary Fig. S14c.

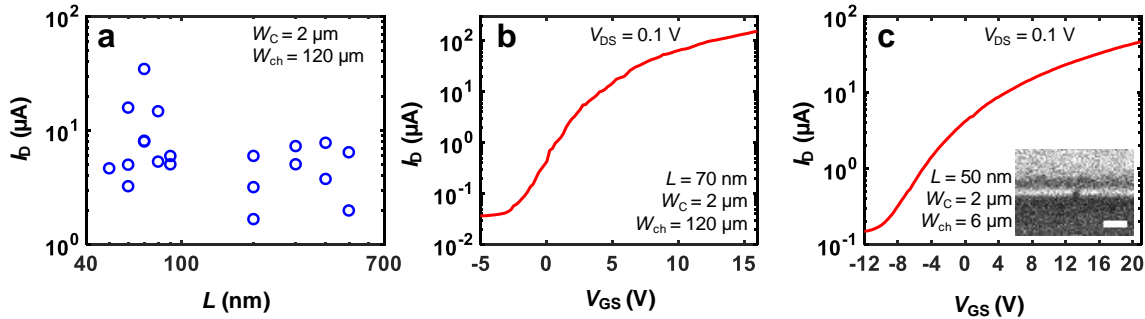

**Supplementary Fig. S14 | MoS₂ *Type A* transistors**. **a**, Drain current $I_D$ vs. channel length at a drain-source voltage $V_{DS} = 0.1$ V and an overdrive voltage $V_{GS} - V_T = 8$ V. **b**, Measured $I_D$ vs. $V_{GS}$ of the device with the highest on-current (also see Fig. 3f). **c**, Electrical characteristic of a *Type A* flexible MoS₂ field-effect transistor with the shortest channel length of 50 nm. Inset shows a scanning-electron microscopy cross-section of such a 50 nm channel. Scale bar: 200 nm.

In Supplementary Fig. S15 we display the histogram for $V_T$ of all MoS₂ devices measured in this work. Academic fabrication and growth variations are the leading cause for the $V_T$ variability, which could be much improved with industrial process optimization (beyond the scope of this work).

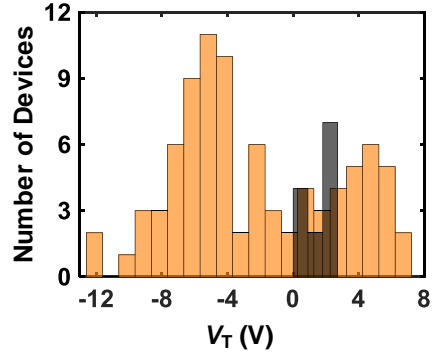

**Supplementary Fig. S15 | Threshold voltage ($V_T$) variability of all flexible MoS$_2$ FETs.** $V_T$ extracted with the linear extrapolation method. Most data are for EOT ~16.4 nm (orange) and some data for EOT ~13.7 nm (black). We measured ~100 *Type A* and *Type B* devices, and the geometry, process flows or channel length did not discernably correlate with any $V_T$ changes.

## J. Drain current vs. channel length in flexible MoS$_2$ FETs and Modeling

The drain current $I_D$ that is obtained when probing a short-channel FET typically has contributions from the channel resistance $R_{ch}$ and contact resistance for $R_C$, which makes the accurate extraction of the intrinsic mobility $\mu_{FE}$ difficult. Furthermore, since short-channel devices can be contact-dominated, an estimation of $R_C$ is also important. Taking into account $R_{ch}$ and equal source and drain $R_C$ we can estimate the overall resistance $R_{tot}$ (all normalized by device width $W$, in units Ω·m) as follows:

$$R_{tot}/W = 2R_C/W + R_{ch}/W.$$

Here, this is applied to *Type B* devices ($W = W_{ch} < W_C$), whereas *Type A* devices need an addition correction for current spreading (Section K below). In the linear transistor operating region (at small $V_{DS}$), $R_{ch}$ can be approximated as

$$R_{ch} = \frac{V_{DS}}{I_D} W \approx \frac{L}{\mu_{FE} C_{ox} (V_{GS} - V_T)},$$

where $I_D$, $L$, $\mu_{FE}$, $C_{ox}$, $V_{GS}$ and $V_T$ are the drain current, transistor channel length, intrinsic field-effect mobility, dielectric capacitance, gate-source voltage and threshold voltage, respectively. Hence, taking into account $R_{tot}$ when measuring the drain current ($I_{D,meas}$) while applying a voltage between drain and source ($V_{DS,appl}$), we obtain the following expression for $I_{D,meas}$ (normalized by $W$, unit: A m$^{-1}$):

$$I_{D,meas} = \frac{V_{DS,appl}}{R_{tot}} = \frac{V_{DS,appl}}{2R_C + \frac{L}{\mu_{FE} C_{ox} (V_{GS} - V_T)}}.$$

Thus, we have an expression for $I_{D,meas}$ where the denominator has two competing components which limit the maximum $I_{D,meas}$ that can be obtained. Further, we can find that for ultra-scaled transistors in the limit $L \rightarrow 0$, the maximum $I_{D,meas}$ is fully limited by $R_C$. In contrast, at long channel devices (here, $L \geq 10$ μm) the $R_C$ no longer has significant impact and $I_{D,meas}$ is mainly defined by $\mu_{FE}$, given the electrostatics and carrier concentration are fixed ($C_{ox}$, $V_{GS}$ and $V_T$ constant).

In the following, we use this model to identify lower and upper bounds for $R_C$ and $\mu_{FE}$ based on our experimentally obtained results in flexible MoS$_2$ FETs with channel lengths ranging from 10 μm down to 50 nm. For that we extract $I_{D,meas}$ at $V_{DS} = 0.1$ V (Fig. 3d) at an overdrive gate voltage ($V_{ov} = V_{GS} - V_T = 8$ V) using a $V_T$ extracted from a linear fit of $g_m$ vs $V_{GS}$. In addition, our model needs $C_{ox}$, which in this case is 0.21 μF cm$^{-2}$ or EOT ~16.4 nm (Supplementary Fig. S7a). With that, we can use $R_C$ and $\mu_{FE}$ as fitting parameters. We obtain upper bounds for $R_C = 2.4$ kΩ μm and $\mu_{FE} = 28$ cm$^2$V$^{-1}$s$^{-1}$ for *Type B* devices. Taking into account also the best Type A devices, the upper bounds become $R_C = 0.31$ kΩ μm and $\mu_{FE} = 29$ cm$^2$V$^{-1}$s$^{-1}$ indicating a remarkably reduced $R_C$ for the *Type A* device at $L = 70$ nm. Furthermore, it becomes evident that at channel lengths of 10 μm the devices are dominated by $\mu_{FE}$ as changes in $R_C$ do not significantly impact $I_D$ (dotted blue and solid black lines converge in Fig. 3d). This gives us the opportunity to fix the $\mu_{FE}$ range at $L = 10$ μm and then subsequently fit $R_C$ to the highest data points at the smallest channel lengths, where $R_C$ has a larger impact than $\mu_{FE}$, which gives us an estimate for the best $R_C$. We performed the fitting at $V_{DS} = 0.1$ V (Fig. 3d) to ensure that the devices are in the linear operating regime. We show in Fig. 3c,f that the devices with the highest $I_D$ display effects of self-heating and velocity saturation even below $V_{DS}$ values that would warrant channel pinch-off, which can be commonly observed for sub-100 nm channel length in MoS$_2$ FETs[7,35]. However, this does not impact our model because we only use it at low $V_{DS}$. The model can also be used to predict the extrinsic field-effect mobility $\mu_{FE,ext}$. The above equations can be modified to[36]:

$$\mu_{FE,ext} = \frac{\mu_{FE}}{1 + \frac{\mu_{FE}C_{ox}(V_{GS}-V_T)2R_C}{L}} .$$

We used the same input parameters ($R_C$ and $\mu_{FE}$) for $\mu_{FE,ext}$ as for our $I_D$ fitting. The result shown in Fig. 3e agrees with our extracted $\mu_{FE,ext}$ based on the maximum $g_m$ method, confirming our calculations.

## K. Correction for Lateral Current Spreading in *Type A* FETs

In all our TMD FETs of *Type A*, where the semiconductor width is greater than the electrode width, fringe currents can contribute non-negligibly to the total measured current depending on a number of factors including contact width and spacing, contact resistance and semiconductor mobility. In order to provide an accurate extraction and comparison of $I_D$ and $\mu_{FE,ext}$ for *Type A* devices and estimate the fringe current effects, we define a dimensionless correction factor:

$$\text{CF} = \frac{I_D}{W I_{D,1D}}$$

where $I_D$ is the total current (in μA), $W$ is width of the contact and semiconductor overlap, and $I_{D,1D}$ is the width-normalized current (in μA μm$^{-1}$) in a FET with the same electrical parameters (sheet and contact resistances) but a channel geometry without current spreading. For FETs without a well-defined patterned channel, CF > 1, reflecting the contribution of fringe currents. In the linear regime of the transistor, given $R_C$, $\mu_{FE}$ and bias voltages; $I_{D,1D}$ can be calculated as $V_{DS} \cdot W^{-1} \cdot R_{tot}^{-1}$, with $R_{tot}$ calculated as defined in the previous section. $I_D$ depends on the device and contact geometry as well: we estimate it using 2D finite element method (FEM) simulations using COMSOL Multiphysics to calculate the current distribution. With CF so obtained, width-normalized corrected currents are then

$$I_{D,corr} = \frac{1}{\text{CF}} \frac{I_{D,meas}}{W}.$$

In FEM simulations, the transistor is assumed to be in the linear region of operation with $V_{DS} \ll V_{GS} - V_T$, so the semiconductor sheet resistance is assumed to be the same everywhere (except where the semiconductor overlaps with contacts), and given by

$$R_{sh} = \frac{1}{q \mu_{FE} C_{ox} (V_{GS} - V_T)},$$

where $q$ is the elementary electric charge. For purposes of calculating this current distribution, the contacts are assumed to be edge contacts[84] with a contact resistance per unit width of $R_C$. The edge contact assumption is equivalent to contacts with current transfer length[27] $L_T < 50$ nm.

The two unknown electrical parameters that influence CF are $\mu_{FE}$ and $R_C$. However, for devices without a patterned channel, it is not straightforward to extract these directly from electrical data while simultaneously correcting for fringe currents: a range of CF are possible for different combinations of $\mu_{FE}$ and $R_C$. For the lower end of this range, we assume an $R_C = 250$ Ω μm (best prior reported results for CVD MoS$_2$ with Au contacts)[24], and fit $\mu_{FE}$ to get the measured $I_D$. For the upper CF estimate, we assume $\mu_{FE}$ about 2-fold higher than in our best devices (for CVD MoS$_2$ a similar value to best prior results on silicon)[24] resulting in $\mu_{FE} = 56$ cm$^2$V$^{-1}$s$^{-1}$ for MoS$_2$, 5 cm$^2$V$^{-1}$s$^{-1}$ for MoSe$_2$ and 10 cm$^2$V$^{-1}$s$^{-1}$ for WSe$_2$, then fit $R_C$ to get the measured $I_D$. The range of CF calculated this way yields a range of $I_{D,corr}$ and $\mu_{FE,ext}$, which is illustrated as vertical bars in e.g., Figs. 3d and 3e.

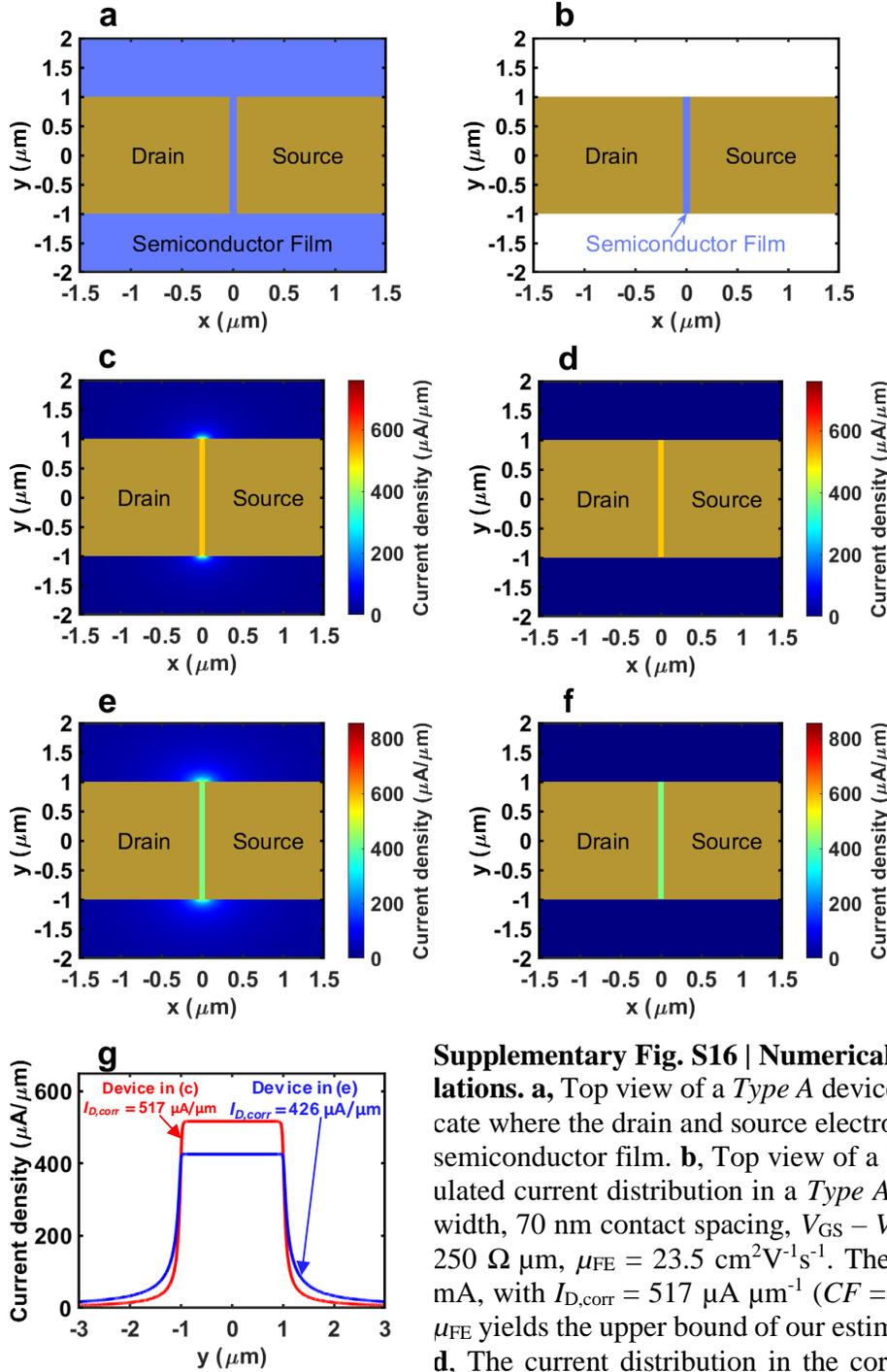

**Supplementary Fig. S16 | Numerical current spreading simulations. a,** Top view of a *Type A* device. The regions in gold indicate where the drain and source electrodes are in contact with the semiconductor film. **b,** Top view of a *Type B* device. **c,** The simulated current distribution in a *Type A* device with 2 μm contact width, 70 nm contact spacing, $V_{GS} - V_T = 10$ V, $V_{DS} = 1$ V, $R_C = 250$ Ω μm, $\mu_{FE} = 23.5$ cm$^2$V$^{-1}$s$^{-1}$. The resulting current is 1.225 mA, with $I_{D,corr} = 517$ μA μm$^{-1}$ ($CF = 1.186$). This set of $R_C$ and $\mu_{FE}$ yields the upper bound of our estimated range of $I_{D,corr}$ values. **d,** The current distribution in the corresponding *Type B* device with the same electrical parameters as in c. The width-normalized current is equal to $I_{D,corr} = 517$ μA μm$^{-1}$. **e,** The simulated current distribution in a *Type A* device with 2 μm contact width, at the same bias conditions as in c but with $R_C = 873$ Ω μm, $\mu_{FE} = 56$ cm$^2$V$^{-1}$s$^{-1}$. The resulting current is also 1.225 mA, with $I_{D,corr} = 426$ μA μm$^{-1}$ ($CF = 1.438$). This set of $R_C$ and $\mu_{FE}$ yields the lower bound of our estimated range of $I_{D,corr}$ values. **f,** The current distribution in the corresponding *Type B* device with the same electrical parameters as in e. The width-normalized current is equal to $I_{D,corr} = 426$ μA μm$^{-1}$. **g,** The current density midway between the contacts ($x = 0$) as a function of the y-coordinate. Red and blue curves correspond to the *Type A* devices c and e, respectively.

An example for the current spreading correction is displayed in Supplementary Fig. S16 for a 70 nm long $MoS_2$ FET. The FEM simulation result displays the current flow paths for two different scenarios of fitted $R_C/\mu_{FE}$ (Supplementary Figs. S16c,e). It is visible that the spreading is more pronounced for the scenario with higher $\mu_{FE}$ and higher $R_C$. In Supplementary Figs. S16d and f, the respective $I_D$ distributions with a hypothetical channel width = electrode width are shown which yield $I_{D,1D}$. For this device, we reported in the main manuscript the average values for these two bounds, which is ~470 $\mu A\,\mu m^{-1}$. Finally, Supplementary Fig. S16g displays the current distribution in the $y$-direction (perpendicular to the channel) at the channel center (defined as $x = 0$), which visualizes the current spreading effect for the two different fitting scenarios.

## L. Benchmarking Tables

| Reference | Synthesis method | Thickness (nm) | Length (nm) | $\mu_{FE}$ or $\mu_{FE,ext}$ ($cm^2V^{-1}s^{-1}$) | $I_D$ at $V_{DS} = 1V$ ($\mu A\,\mu m^{-1}$) |
|---|---|---|---|---|---|
| Kwon et al. [40] | exfoliated | 30-80 | 7000 | 44.8 | 4.1 |
| Chang et al. [41] | exfoliated | 7.9 | 1000 | 30* | 13 |
| Yoon et al. [42] | exfoliated | 3.075 (5 layers) | 4000 | 4.7 | 0.3 |
| Lee et al. [43] | exfoliated | 1.845 (3 layers) | 800 | 29 | NA |
| Salvatore et al. [4] | exfoliated | 3.5 | 4300 | 19 | 1.3 |
| Yoo et al. [44] | exfoliated | 66.5 | 8300 | 83.5 | 3 |
| Song et al. [45] | exfoliated | 79.3 | 22600 | 141.3 | 1.2 |
| Cheng et al. [16] | exfoliated | 1.845 (3 layers) | 116 | NA | 48*[2] |
| Cheng et al. [16] | exfoliated | 1.845 (3 layers) | 68 | NA | 135*[3] |
| Ma et al. [46] | exfoliated | 18.3 | 17500 | 15.5 ±15.9 | 0.32 |
| Chang et al [17] | CVD | 0.615 (1 layer) | 750 | 31* | 66 |
| Amani et al. [47] | CVD | 0.615 (1 layer) | 13000 | 18.9* | 0.2 |
| Shinde et al. [29] | CVD | 0.615 (1 layer) | 4000 | 6.7 ±20 | 2.2 |
| Woo et al. [48] | CVD | 1.538 (2.5 layers) | 10000 | 9 | 1.4 |
| Park et al. [49] | CVD | 1.23 (2 layers) | 4000 | 17.4 | 0.13 |

**Supplementary Table S1:** Literature values for flexible $MoS_2$ field-effect transistors used to generate Figs. 4a,b. The synthesis method denotes whether the material was mechanically exfoliated or grown by chemical vapor deposition (CVD). * indicates that in these cases the field-effect mobility $\mu_{FE}$ was extracted by the y-function method and contact resistance $R_C$ is excluded. For the other cases the method was either not specified or the transconductance $g_m$ maximum method was used which results in an extrinsic $\mu_{FE}$ ($\mu_{FE,ext}$). The drain current $I_D$ is in most cases specified at a drain-source voltage $V_{DS}$ = 1 V, unless labeled: *[2] $V_{DS}$ = 2V, *[3] $V_{DS}$ not specified.

| Reference | Channel material | Length (nm) | on/off ratio | $I_D$ at $V_{DS}$ = 0.5 V ($\mu A\ \mu m^{-1}$) |
|---|---|---|---|---|
| Park et al. [51] | graphene | 140 | 1.5 | 248 |
| Yeh et al. [52] | graphene | 200 | 8.8 | 516 |
| Zhai et al. [50] | single-crystal silicon (c-Si) | 150 | $6 \times 10^7$ | 369 |
| Shahrjerdi et al. [39] | single-crystal silicon (c-Si) | 30 | $2 \times 10^5$ | 714 |
| Wang et al. [53] | InSnO (ITO) | 160 | $7 \times 10^8$ | 34 |
| Münzenrieder et al. [54] | InGaZnO (IGZO) | 160 | 7.1 | 119 |
| Cheng et al. [16] | $MoS_2$ | 116 | $2 \times 10^6$ | 48* |
| Cheng et al. [16] | $MoS_2$ | 68 | $10^6$ | 135*² |

**Supplementary Table S2:** Literature values for flexible field-effect transistors with channel lengths $\leq$ 200 nm used to generate Fig. 4c. The drain current $I_D$ is in most cases specified at a drain-source voltage $V_{DS}$ = 0.5 V, unless labelled: * $V_{DS}$ = 2V, *² $V_{DS}$ not specified.

## Additional References